# A Deep Reinforcement Learning Framework for Dynamic Portfolio Optimization: Evidence from China's Stock Market

Gang Huang[1]*, Xiaohua Zhou[1†] and Qingyang Song[2†]
[1]*Department of Applied Economics, Chongqing University, 174 Shazheng Street, Chongqing, 400044, China.
[2]Department of Technology Economics and Management, Chongqing University, 174 Shazheng Street, Chongqing, 400044, China.

*Corresponding author(s). E-mail(s): huanggangvoyager@163.com;
Contributing authors: zhxiaoh@aliyun.com; song211@126.com;
[†]These authors contributed equally to this work.

**Abstract**

Artificial intelligence is transforming financial investment decision-making frameworks, with deep reinforcement learning demonstrating substantial potential in robo-advisory applications. This paper addresses the limitations of traditional portfolio optimization methods in dynamic asset weight adjustment through the development of a deep reinforcement learning-based dynamic optimization model grounded in practical trading processes. The research advances two key innovations: first, the introduction of a novel Sharpe ratio reward function engineered for Actor-Critic deep reinforcement learning algorithms, which ensures stable convergence during training while consistently achieving positive average Sharpe ratios; second, the development of an innovative comprehensive approach to portfolio optimization utilizing deep reinforcement learning, which significantly enhances model optimization capability through the integration of random sampling strategies during training with image-based deep neural network architectures for multi-dimensional financial time series data processing, average Sharpe ratio reward functions, and deep reinforcement learning algorithms. The empirical analysis validates the model using randomly selected constituent stocks from the CSI 300 Index, benchmarking against established financial econometric optimization models. Backtesting results demonstrate the model's efficacy in optimizing portfolio allocation and mitigating investment risk, yielding superior comprehensive performance metrics.

**Keywords**：Portfolio Management; Decision Optimization; Dynamic Optimization; Deep Reinforcement Learning

## 1 INTRODUCTION

In recent years, Artificial Intelligence (AI) has achieved significant technological advances, notably in natural language processing. ChatGPT, developed by OpenAI, has catalyzed extensive discourse on AI's potential through its exceptional language comprehension and generation capabilities. The system's success is predominantly attributed to "Reinforcement Learning from Human Feedback" (RLHF), an innovative methodology that substantially enhances AI system performance and alignment through the integration of human feedback into the reinforcement learning process. The technological foundation of RLHF is Deep Reinforcement Learning (DRL), an advanced machine learning paradigm that synthesizes deep learning and reinforcement learning methodologies. While DRL has demonstrated remarkable efficacy in natural language processing and exhibited substantial potential across domains including game AI and robotic control, its applications in finance remain predominantly exploratory, particularly in the complex domain of portfolio optimization.

Portfolio optimization constitutes a fundamental challenge in finance, focusing on the systematic allocation of funds across multiple assets based on investment decisions, manifesting through dynamic adjustments in portfolio asset weights. Traditional portfolio optimization methodologies, originating from Modern Portfolio Theory[1] and evolving through subsequent enhancements, have contributed substantially to the field's development. However, these approaches present inherent limitations, including restrictive assumptions regarding asset return distributions, subjective utility function specifications, and insufficient adaptability to dynamic market conditions.

This research examines the implementation potential of deep reinforcement learning in portfolio optimization through the development of novel reward functions and deep neural network architectures, aimed at constructing an intelligent model for effective dynamic asset allocation. The study advances both theoretical and practical contributions by introducing innovative approaches to portfolio optimization while establishing new trajectories for artificial intelligence applications in financial domains.

## 2 LITERATURE REVIEW

Markowitz[1] established modern portfolio theory, pioneering the application of quantitative analysis methods in portfolio optimization. Samuelson [2] contended that Markowitz's model addressed single-period investment problems but was inadequate for multi-period asset allocation, consequently proposing a utility function for analyzing wealth planning problems. Subsequent researchers, including Kelly [3], Merton [4], and numerous scholars in behavioral finance, extended the application of utility functions in asset allocation optimization. However, utility function-based optimization approaches exhibit inherent limitations, particularly in the inherent subjectivity of function selection and the unverified universal applicability of chosen functions. The Black-Litterman (BL) model[5,6] represents another approach incorporating subjective elements, proposing that markets possess an implicit equilibrium return, where asset returns under equilibrium allocation serve as prior returns. In this model, expected returns represent a weighted average of prior returns and investors' subjective expectations. However, the significant subjectivity in establishing confidence levels for investors' subjective expectations has resulted in the absence of a unified standard for measuring the equilibrium return rate of portfolio assets.

Beyond traditional financial econometric analysis methods, operations researchers Charnes et al.[7] introduced Data Envelopment Analysis (DEA), a non-parametric analytical framework for asset allocation optimization. Subsequently, Kirkpatrick[8] integrated the simulated annealing algorithm into portfolio optimization, based on principles derived from natural sciences. In parallel, Arnone et al.[9] implemented genetic algorithms for portfolio selection to minimize investment risk. However, these models universally treat the portfolio weight adjustment process as static, neglecting the temporal dimension and failing to incorporate how asset allocations evolve in response to the sequential nature of trading activities.

Furthermore, classical asset allocation models, including Markowitz's framework, compute portfolio returns by multiplying contemporaneous asset weights with Expected Returns to derive the period's portfolio return. This is expressed as $R = \sum_{i=1}^{n} \omega_i \gamma_i$, where R represents the portfolio return, i denotes the number of assets, $\omega_i$ signifies the asset weight, and $\gamma_i$ represents the corresponding Expected Return. However, in a realistic dynamic trading environment, the portfolio's terminal return should be computed by multiplying the previous period's asset weights with the subsequent period's asset returns, expressed as $R_t = \sum_{i=1}^{n} \omega_{i,t-1} \gamma_{i,t}$ , where $\gamma_{i,t}$ represents the realized return of asset i in period t (not the Expected Return), and $\omega_{i,t-1}$ represents the portfolio's asset weight allocation in period t-1. This fundamental distinction can lead to significant discrepancies - modeling errors in the trading process inevitably compromise the practical efficacy of these models. Notably, numerous widely-implemented optimization models in finance, including the Conditional Value at Risk model[10], Risk Parity model[11], and Hierarchical Risk Parity model[12], entirely disregard the temporal evolution of asset weights. Consequently, neither conventional financial econometric analysis methods nor sophisticated approaches such as DEA, simulated annealing algorithms, and genetic algorithms can adequately capture the dynamic nature of portfolio weight adjustments during the trading process, thereby failing to achieve optimal asset allocation strategies.

Deep Reinforcement Learning (DRL) represents a dynamic modeling paradigm. The "Deep" denomination in DRL derives from its incorporation of deep neural networks, which supersede the conventional artificial neural networks employed in early Reinforcement Learning (RL), including fully connected and recurrent neural networks. This architectural enhancement has substantially improved RL's capacity for objective function approximation. Early applications of RL in asset management primarily employed Policy Gradient (PG)[13,14] and Q-learning algorithms. Moody[15] introduced a single-asset management model utilizing the PG algorithm, with subsequent derivative models predominantly focusing on single-risk asset management or fixed investment decision frameworks, as exemplified by Dempster et al.'s[16] automated forex trading model and Zhang et al.'s[17] asset management framework. In parallel, Ralph Neuneier[18], Gao et al.[19], and Lee et al.[20] implemented Q-learning algorithms for asset management, though these models remained confined to single-asset management. Furthermore, notable contributions to the research field of DRL applications in single-asset trading have been made by Wu et al.[21], Liu et al.[22], Pourahmadi et al.[23], and Kochliaridis et al.[24], among others. However, some scholars have neglected the design of deep neural networks when applying DRL to optimize asset allocation, such as Wang et al.[25], while others have overlooked the allocation of asset weights and missed the basic constraint condition that asset weights sum to 1 ($\sum\omega_{i,t} = 1$), such as Jiang et al.[26].

Recent advancements in computational capabilities and dynamic optimization theory have led to the widespread adoption of DRL in portfolio asset management research. Jiang et al.[27] proposed a DRL portfolio management model for asset optimization in the cryptocurrency market. The model incorporated the definitions of relative price vectors and transaction costs from Ormos et al.[28]. However, Ormos et al. misinterpreted the dynamic changes of assets in their paper, resulting in incorrect transaction cost derivations. Due to the adoption of the same derivation methodology, Jiang et al.[27]'s derivation of transaction cost rates exhibited comparable mathematical inconsistencies. While Jiang et al.[27] subsequently provided correct implementation formulas through approximation methods, the model's effectiveness in alternative capital markets requires further validation[29].

Under short-selling restrictions (long-only positions), the reward function in current DRL portfolio weight optimization models primarily consists of portfolio returns[30]. However, DRL models using this reward function have not performed well in Chinese stock market[29], leading Qi Yue et al.[31] to artificially set fixed investment weights to achieve satisfactory backtesting results. This approach, however, contradicts the original intention of using DRL models for automatic asset weight optimization.

In the field of DRL portfolio applications, researchers have demonstrated that implementing reward functions to enhance DRL's asset optimization performance represents an effective approach. Multiple scholars have developed new reward functions to improve DRL's portfolio optimization performance: Wu et al.[32] investigated Taiwan stock market portfolios using a customized Sharpe ratio reward function (Annual Return/Annualized Standard Deviation of Return). However, their research did not specify the underlying RL algorithm implemented. Almahdi et al.[33] incorporated the Calmar ratio as the optimization objective in the reward function, integrating it with Recurrent Reinforcement Learning (RRL), a derivative algorithm of PG, to optimize US stocks and emerging market assets. Aboussalah et al.[34] developed a Sharpe ratio reward function compatible with RRL derivative algorithms for asset allocation, though this reward function is fundamentally equivalent to the Sharpe ratio reward function of the PG algorithm. Furthermore, Lim et al[35] employ a reward function based on the Net Asset Value of the portfolio to develop an RL-based strategy for dynamic portfolio rebalancing that optimizes investment performance under varying market conditions. A comprehensive review of existing literature reveals that no research has established appropriate Sharpe ratio reward functions specifically designed for the algorithmic characteristics of Actor-Critic.

A critical review of existing DRL portfolio optimization literature reveals a significant limitation: most studies fail to demonstrate the convergence of their reward functions during experimentation[36, 37, 38, 39, 40]. Even in cases where reward function convergence is presented, the convergence exhibits substantial deficiencies. For example, Yang[41]'s TC-MAC algorithm, which employs GNN to capture dynamic relationships between assets, demonstrates concerning convergence behavior: despite an initial capital of $10,000, the reward function converges to merely $3,000, indicating potentially significant losses. Similarly, Sun et al.[42], in their integration of DRL with the Black-Litterman model, artificially constrain portfolio weights through predetermined long-short position limits, fundamentally contradicting DRL's autonomous optimization capability. These cases exemplify fundamental limitations in current approaches, raising concerns about the models' stability and reliability in practical applications. This methodological gap undermines the robustness of existing DRL-based portfolio optimization frameworks and calls into question their practical applicability in real-world investment scenarios.

This paper implements DRL methodology based on artificial intelligence principles to optimize portfolio asset weights, effectively eliminating subjective bias in model implementation while comprehensively addressing the dynamic characteristics of asset weight variations in real-world trading environments. The research presents two primary innovations:

First, we introduce a novel Sharpe ratio reward function specifically engineered for Actor-Critic DRL algorithm characteristics. While seminal research in RL asset management applications by Moody[15] and Gao[19] employed the Sharpe ratio as a reward function, their designs were constrained to simple structures of PG algorithm and Q-learning algorithm. Similarly, the Sharpe ratio reward function developed by Aboussalah et al.[34] did not adequately address the specific requirements of Actor-Critic algorithms. Our proposed reward function incorporates the architectural characteristics of Actor-Critic systems, implementing step-size normalization to enhance model stability and optimize the guidance of portfolio dynamic optimization processes. Notably, our reward function design ensures stable convergence during model training and consistently achieves positive average Sharpe ratios, addressing a critical gap in current DRL portfolio optimization applications where reward function convergence remains a significant challenge.

Second, this research introduces an innovative comprehensive methodology for portfolio optimization utilizing deep reinforcement learning. This approach significantly enhances model optimization capability through the integration of random sampling strategies during training with image-based deep neural network architectures for multi-dimensional financial time series data processing, average Sharpe ratio reward functions, and deep reinforcement learning algorithms. Specifically, the network architecture incorporates VGG network design principles from computer vision to construct a deep neural network framework for processing three-dimensional time series data. Additionally, the model's generalization capability is enhanced and overfitting risk is minimized through the random selection of continuous trading data from the dataset during training, establishing a robust technical framework for effective dynamic portfolio optimization.

This research implements long-only position constraints and applies the proposed

DRL model to optimize portfolios comprising CSI300 constituent stocks. The optimization outcomes are systematically benchmarked against multiple econometric optimization models to evaluate the DRL model's efficacy in asset allocation optimization. The research methodology adheres rigorously to the DRL model's modeling framework, establishing comprehensive guidelines for future research endeavors. The significance of this study extends across both theoretical and practical domains: it introduces a novel portfolio optimization methodology to the academic literature while providing an effective solution for portfolio management practitioners. The model systematically incorporates the dynamic characteristics of asset weight variations in real trading environments, demonstrating significant potential for enhanced performance in practical applications.

## 3 DRL MODEL CONFIGURATION

Deep Reinforcement Learning (DRL) represents a dynamic optimization method conforming to the Markov Decision Process (MDP) framework. The portfolio trading process can be conceptualized as an MDP, where the trajectory from account initiation to trading completion is represented by $\tau = (S_0, A_0, R_1, S_1, A_1, R_2, S_2, A_2, R_3, \cdots)$, constituting an episode. This framework enables the application of DRL theory for modeling the trading process. Following the DRL modeling framework, this study defines a portfolio trader as an agent, establishes the state (environment), action, and reward specifications, and implements a DRL algorithm with deep neural networks for portfolio optimization.

### 3.1 State Space Configuration

The state space in DRL constitutes the environment for agent interaction. Following the Efficient Market Hypothesis, all information affecting asset values is embedded in asset prices; consequently, the state space is constructed exclusively using daily asset price data. This study adopts the three-dimensional state space configuration proposed by Jiang et al.[27] in modeling the DRL environment, based on two fundamental considerations. First, DRL achieved its breakthrough in artificial intelligence through video game applications[43]. Video game displays comprise three-dimensional data structures, which are inherently suitable for deep neural network processing. Deep neural networks had previously demonstrated exceptional progress in image recognition, achieving human-comparable performance in this domain. Second, traditional financial econometric models typically employ dimensionality reduction techniques, such as Principal Component Analysis (PCA), to reduce analytical complexity. However, these methods frequently result in significant loss of valuable information, with the loss magnitude increasing proportionally with the number of data features. In contrast, deep neural networks possess superior nonlinear function approximation capabilities, enabling effective analysis of complex feature interrelationships and addressing the limitations of traditional financial econometric models. Consequently, this study implements a three-dimensional temporal data structure for the state space, effectively leveraging the advanced data processing capabilities of deep neural networks.

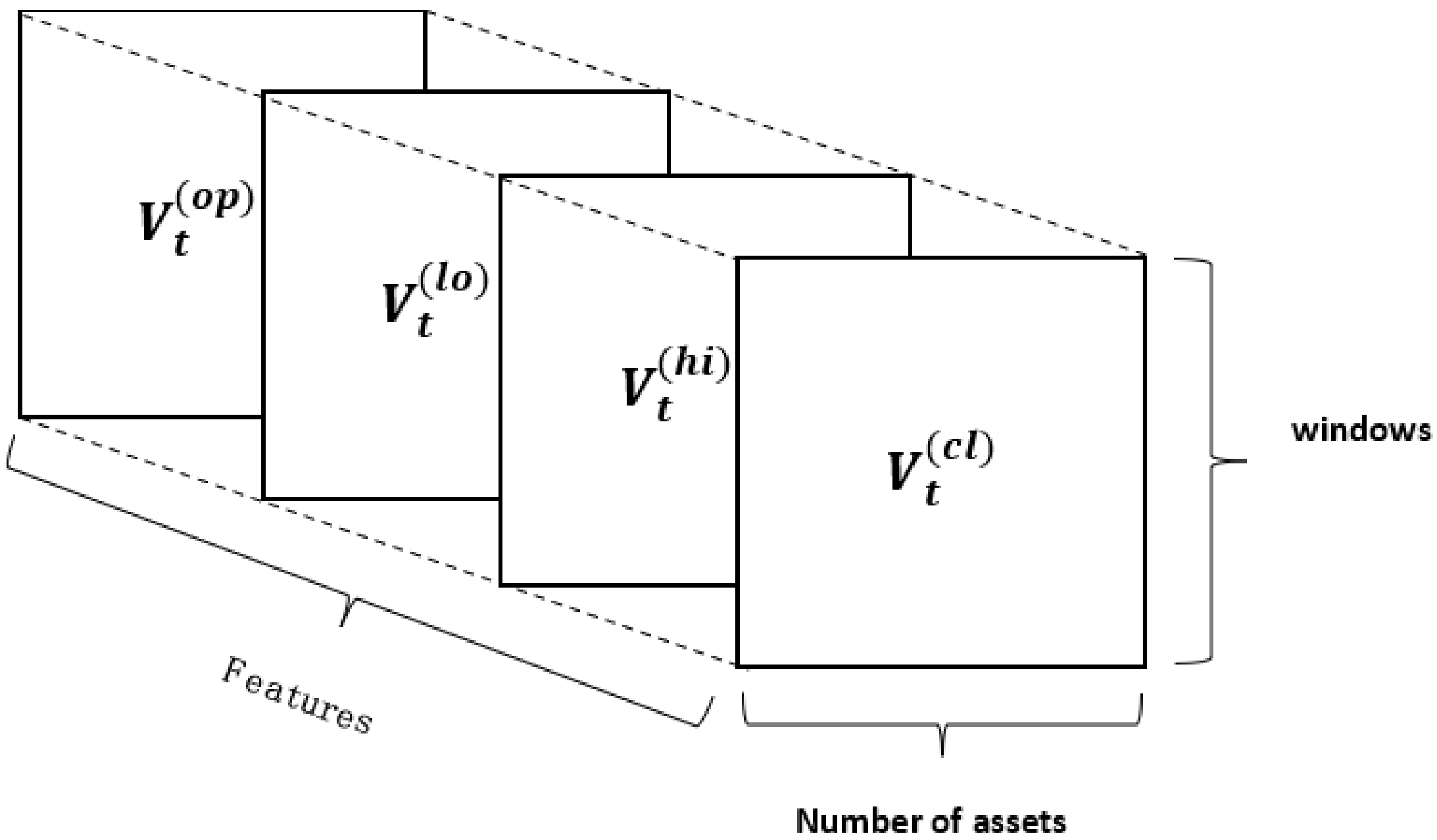

Fig1 Data structure of state $X_t$

The state is defined as $S_t = X_t$, where the price tensor $X_t$ comprises four data features: daily opening price $V_t^{(op)}$, lowest price $V_t^{(lo)}$, highest price $V_t^{(hi)}$, and closing price $V_t^{(cl)}$. The data structure is illustrated in Figure 1, with the tensor $X_t$ calculation formula given by:

$$\begin{aligned}
\boldsymbol{V}_{\boldsymbol{t}}^{(\boldsymbol{op})} &= [\boldsymbol{v}_{\boldsymbol{t-n+1}}^{(\boldsymbol{op})} \oslash \boldsymbol{v}_{\boldsymbol{t}} | \boldsymbol{v}_{\boldsymbol{t-n+2}}^{(\boldsymbol{op})} \oslash \boldsymbol{v}_{\boldsymbol{t}} | \cdots | \boldsymbol{v}_{\boldsymbol{t-1}}^{(\boldsymbol{op})} \oslash \boldsymbol{v}_{\boldsymbol{t}} | \boldsymbol{v}_{\boldsymbol{t}}^{(\boldsymbol{op})} \oslash \boldsymbol{v}_{\boldsymbol{t}}] \\
\boldsymbol{V}_{\boldsymbol{t}}^{(\boldsymbol{lo})} &= [\boldsymbol{v}_{\boldsymbol{t-n+1}}^{(\boldsymbol{lo})} \oslash \boldsymbol{v}_{\boldsymbol{t}} | \boldsymbol{v}_{\boldsymbol{t-n+2}}^{(\boldsymbol{lo})} \oslash \boldsymbol{v}_{\boldsymbol{t}} | \cdots | \boldsymbol{v}_{\boldsymbol{t-1}}^{(\boldsymbol{lo})} \oslash \boldsymbol{v}_{\boldsymbol{t}} | \boldsymbol{v}_{\boldsymbol{t}}^{(\boldsymbol{lo})} \oslash \boldsymbol{v}_{\boldsymbol{t}}] \\
\boldsymbol{V}_{\boldsymbol{t}}^{(\boldsymbol{hi})} &= [\boldsymbol{v}_{\boldsymbol{t-n+1}}^{(\boldsymbol{hi})} \oslash \boldsymbol{v}_{\boldsymbol{t}} | \boldsymbol{v}_{\boldsymbol{t-n+2}}^{(\boldsymbol{hi})} \oslash \boldsymbol{v}_{\boldsymbol{t}} | \cdots | \boldsymbol{v}_{\boldsymbol{t-1}}^{(\boldsymbol{hi})} \oslash \boldsymbol{v}_{\boldsymbol{t}} | \boldsymbol{v}_{\boldsymbol{t}}^{(\boldsymbol{hi})} \oslash \boldsymbol{v}_{\boldsymbol{t}}] \\
\boldsymbol{V}_{\boldsymbol{t}}^{(\boldsymbol{cl})} &= [\boldsymbol{v}_{\boldsymbol{t-n+1}} \oslash \boldsymbol{v}_{\boldsymbol{t}} | \boldsymbol{v}_{\boldsymbol{t-n+2}} \oslash \boldsymbol{v}_{\boldsymbol{t}} | \cdots | \boldsymbol{v}_{\boldsymbol{t-1}} \oslash \boldsymbol{v}_{\boldsymbol{t}} | \boldsymbol{1}]
\end{aligned} \tag{1}$$

Here, $\boldsymbol{v_t}$denotes the closing price vector of assets on trading day *t*, and the symbol $\oslash$ represents element-wise division, where each vector element is divided by its counterpart at the corresponding position. Each element in the price tensor $\boldsymbol{X_t}$ is normalized through division by the closing price vector $\boldsymbol{v_t}$. The window length (windows) specifies the temporal span of observable data for the agent's trading decisions, with each feature layer containing the corresponding features of all risky assets in the portfolio (i.e., assets_num in Figure 1).

### 3.2 Action Space Configuration

The model only considers long positions without short selling. The portfolio weights (i.e., the ratio of asset value to total assets) represent the model's action vector:

$$\boldsymbol{W_t} = \left(\boldsymbol{\omega_{0,t}}, \boldsymbol{\omega_{1,t}}, \boldsymbol{\omega_{2,t}}, \cdots, \boldsymbol{\omega_{m,t}}\right) \tag{2}$$

where $\boldsymbol{\omega_{0,t}}$ represents the weight of the risk-free asset, specifically defined as the cash asset weight in this study. At time *t*, the portfolio weights satisfy the following constraint:

$$\sum_{i=0}^{m} \omega_{i,t} = 1 \tag{3}$$

Under the long-only constraint, $\omega_{i,t} \geq 0$. The portfolio is initialized with exclusively cash assets, characterized by the initial weight vector $\boldsymbol{W_0} = (1,0,\cdots,0)^{\mathrm{T}}$.

### 3.3 Other Elements Derivation and Reward Function Setting

Let vector $\mathbf{P_t}$ denote the closing prices of assets in the portfolio at period *t*, and $\mathbf{Y_t}$ denote the relative price vector:

$$\boldsymbol{Y_t} \triangleq \boldsymbol{P_t} \oslash \boldsymbol{P_{t-1}} = \left(1, p_{1,t} / p_{1,t-1}, \cdots, p_{i,t} / p_{i,t-1}\right)^{\mathrm{T}} \tag{4}$$

Let $C_t$ denote the transaction cost rate of the entire portfolio in period t. The portfolio price $\rho_t$ is expressed as:

$$\rho_t = \rho_{t-1}(1 - C_t)\exp[(\mathbf{ln}\boldsymbol{Y_t}) \cdot \boldsymbol{W_{t-1}}] \tag{5}$$

The daily logarithmic return rate $\gamma_t$ of the portfolio is defined as:

$$\gamma_t = \ln(\rho_t / \rho_{t-1}) \tag{6}$$

The mean $\bar{R}$ and standard deviation $Std(\gamma_t)$ of daily logarithmic return rates are calculated as:

$$\bar{R} = \frac{1}{t_n}\sum_{t=1}^{t_n} \gamma_t \tag{7}$$

$$std(\gamma_t) = \sqrt{\sum_{t=1}^{t_n}(\gamma_t - \bar{R})^2 \big/ t_n} \tag{8}$$

In formulas (7) and (8), $t_n$ denotes the nth trading period, and $\gamma_t$ is derived from the closing price at the end of period t. At market entry, investors hold exclusively cash assets. With the initial trading time point defined as $t = 0$ and $\gamma_0 = 0$, and considering no allocation to risky assets at this point, formulas (5) and (6) indicate that $\gamma_1 = 0$. Consequently, the return-generating period initiates at t=2.

The reward function employs the average annualized Sharpe ratio:

$$\textbf{reward:}\ An_AVGSharpe_t = \sqrt{Freq} \cdot (\bar{R} - r_f) \big/ Steps \cdot std(\gamma_t - r_f) \tag{9}$$

Freq denotes the annual trading days, set at 252 in this paper. $r_f$ represents the risk-free rate, set to 0 for cash assets. Steps denotes the step length in a training episode, with one trading decision per step. The model's training objective consists of maximizing this reward function.

$C_t$ is determined by:

$$C_t = \mu_t \left( \sum_{i=1}^{m} \left| \boldsymbol{\omega}'_{i,t} - \omega_{i,t} \right| \right) \tag{10}$$

$\mu_t$ represents the transaction cost rate per asset in period t, set at $\mu_t = 0.0025$ in this study, constituting a substantially high rate. $\boldsymbol{\omega}'_{i,t}$ denotes a component of weight vector $\boldsymbol{W}'_t$, given by:

$$\boldsymbol{W}'_t = (\boldsymbol{Y_t} \odot \boldsymbol{W_{t-1}}) \,/\, (\boldsymbol{Y_t} \cdot \boldsymbol{W_{t-1}}) \tag{11}$$

where $\odot$ denotes the Hadamard product and $\cdot$ denotes the inner product. $\boldsymbol{W}'_t$ represents the weight values resulting from autonomous price movements between post-trading at *t-1* and pre-trading at *t*, as illustrated in Figure 2 below:

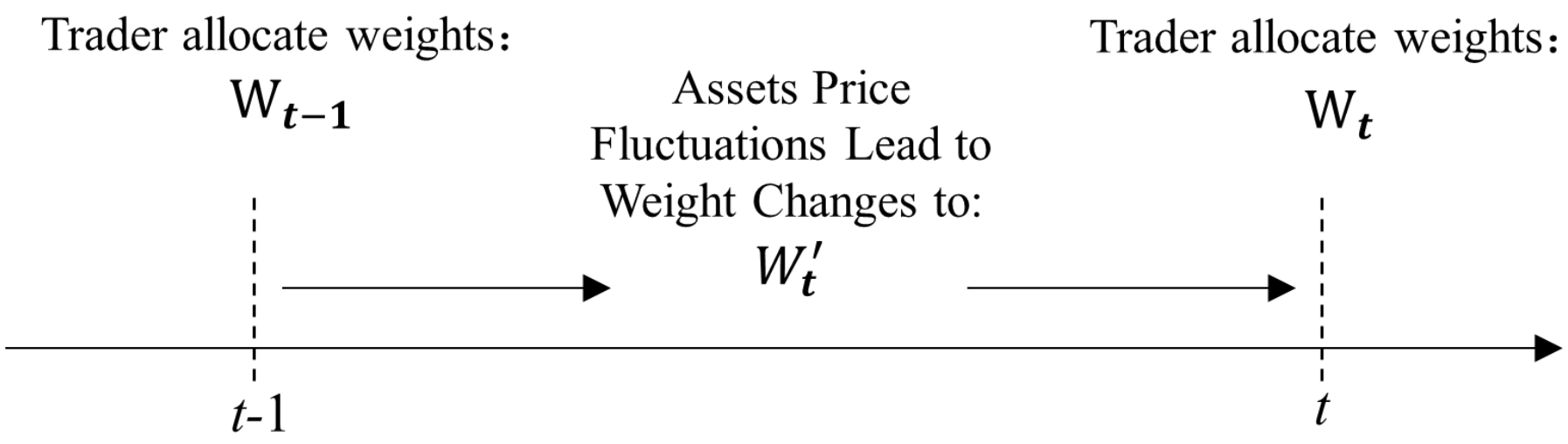

**Figure 2 Changes in asset weights**

# 4 DRL ALGORITHM SELECTION AND NETWORK STRUCTURE

## 4.1 Design of Average Sharpe Ratio Reward Function for Actor-Critic Architecture

In Deep Reinforcement Learning (DRL), algorithms serve as strategic frameworks enabling agents to explore environments and maximize returns through optimal action selection in the action space and reward acquisition in the state space. DRL algorithms comprise two primary categories: on-policy and off-policy approaches, each exhibiting distinct performance characteristics across various tasks. Through comprehensive experimental evaluation of multiple algorithmic structures, we determined that off-policy algorithms demand greater computational resources and demonstrate slower convergence rates. Given hardware constraints, we selected Proximal Policy Optimization (PPO), an on-policy algorithm.

PPO incorporates multiple performance optimization techniques, including Generalized Advantage Estimation (GAE) and value function clipping, fundamentally extending Trust Region Policy Optimization (TRPO)[44], which itself enhances PG algorithms[13,14]. PG algorithms implement episode-based update mechanisms, parallel to policy iteration, optimizing through complete episode sampling. PPO's Actor-Critic architecture uniquely combines both episode-level updates and step-wise updates within episodes. The original PPO paper[45] elegantly presents the algorithm through concise pseudocode utilizing two nested for-loops: an outer loop managing episode updates and an inner loop executing step-wise updates.

Capitalizing on the distinct update mechanism of Actor-Critic algorithms, we developed an innovative average Sharpe ratio reward function calculation method optimized for Actor-Critic frameworks, illustrated through PPO implementation. The methodology initializes an empty list R for storing returns from each trading step. During episode execution, the Actor network generates portfolio weight w1, prompting the environment to return price change information y1 (the relative price vector). These parameters, combined with transaction cost c1, determine the portfolio value change p1 at each time step. The value change converts to logarithmic return r1 and appends to returns list R, enabling Sharpe ratio computation at each trading step using the cumulative returns. The detailed implementation methodology is outlined in Definition 1.

PPO extends the episode update mechanism from PG algorithms while implementing step-wise updates within episodes. For optimizing agent performance during training, we normalize the annualized Sharpe ratio by the number of steps in episode updates (formula 9), computing the agent's average Sharpe ratio at each temporal point per episode. This methodology ensures reward comparability across varied episode lengths and trading sequences, substantially enhancing model training stability.

---

**Definition 1**. Average Sharpe Ratio Reward Function for PPO(Actor-Critic) Algorithm

```
Environment Parameters:
    steps: total number of steps T in an episode
    w0: portfolio weight vector at previous timestep
    p0: portfolio value at previous timestep

Environment Variables:
    R = []    # Initialize cumulative returns list for Sharpe ratio calculation

Reward Definition:
    for each episode do:
        for t = 1 to T do:
            1. State-Action Interaction:
                w1 = πθ(st)
                y1 = price relative vector from environment

            2. Portfolio Value Update:
                compute transaction cost c1
                p1 = p0 * (1 - c1) * exp(dot(log(y1), w0))

            3. Return Calculation:
                r1 = log(p1 / p0)
                append r1 to R

            4. Reward Function:
                reward = sharpe(R) / T
```

5. State Update:

store (st, w1, r1, st+1, reward)

w0 ← w1

p0 ← p1

end for

end for

Output: Timestep reward signals for DRL training

The implementation leverages the Stable-Baselines3 (SB3) framework for PPO deployment, incorporating the innovative average Sharpe ratio reward function within the environment state for portfolio optimization. Crucially, the environment state maintains independence from the DRL algorithm, with the reward function implementation residing in the environment without modifying the core PPO algorithm logic in SB3. Empirical results demonstrate that this average Sharpe ratio reward function effectively harnesses PPO (Actor-Critic) algorithm characteristics, yielding significant improvements in out-of-sample performance.

### 4.2 Neural Network Design

Early artificial neural networks encountered limitations in developing data-driven theoretical models due to challenges in balancing function approximation accuracy and gradient stability while increasing network depth. Advances in deep neural networks enabled reinforcement learning (RL) algorithm progression[43], facilitating deep reinforcement learning (DRL) development. Deep neural network architecture constitutes a critical component in DRL, with empirical evidence indicating that an efficient network design enhances DRL performance.

Given the three-dimensional state space (i.e., price tensor), we conducted comprehensive experiments comparing several mainstream deep neural architectures, including ResNet[46], VGG[47], and Vision Transformer (ViT)[48]. While ResNet demonstrated superior performance in both training and backtesting phases, its computational overhead and slower training speed posed practical limitations. After careful consideration of the trade-off between model performance and computational efficiency, we implemented the VGG architecture as our baseline network structure, as illustrated in Figure 3:

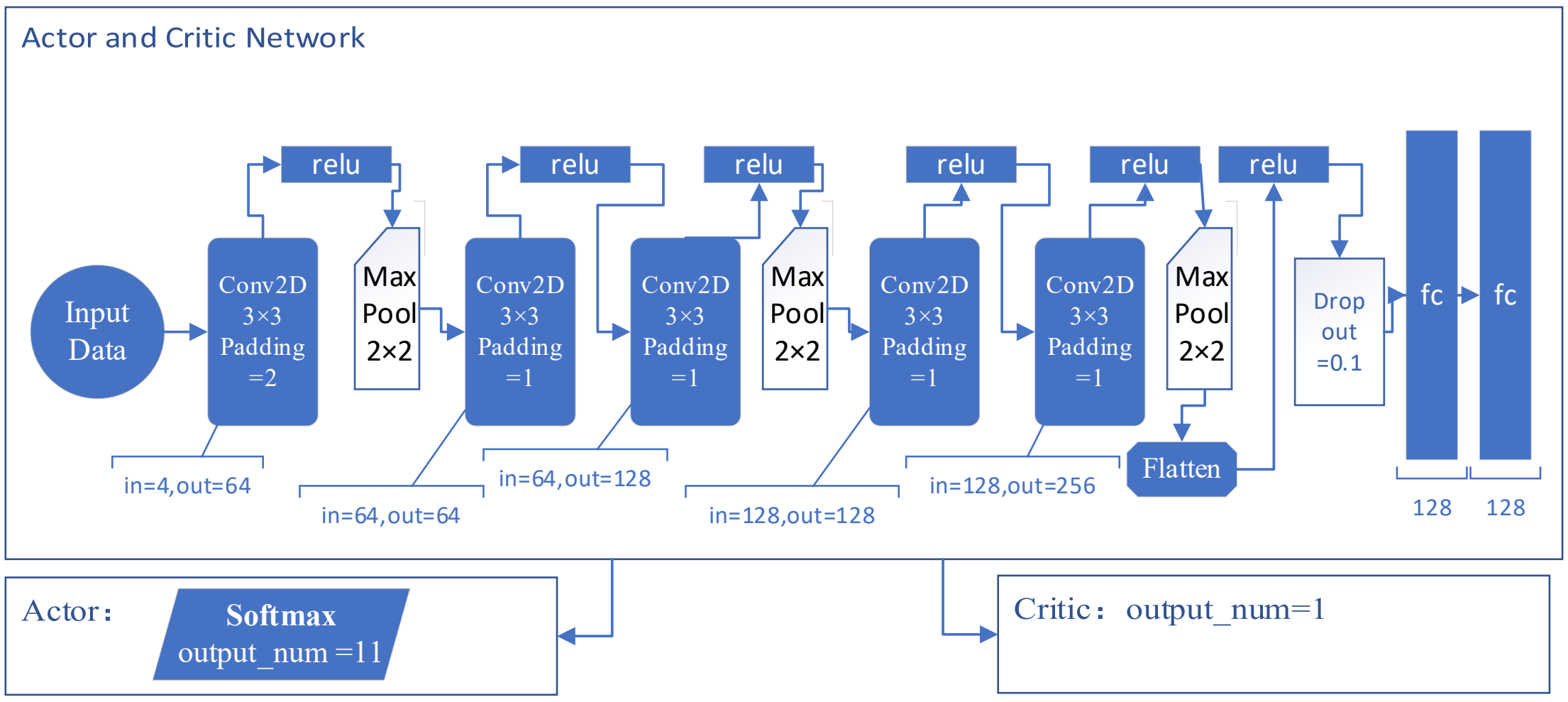

Figure 3 Neural Network Structure

In Figure 3, 'in' specifies the input channel count, while 'out' defines the output channel count. The network architecture incorporates 5 convolutional layers, each employing 3×3 convolution kernels for feature extraction, complemented by Max Pool layers for feature pooling. Post max pooling at the final convolutional layer, a Flatten operation transforms the feature data into a one-dimensional vector, followed by two fully connected (fc) layers executing linear processing, each containing 128 neurons. The Actor network concludes with a softmax activation function, generating the action vector for asset weights, while the Critic network produces the value function output without activation functions. The Actor network's softmax output structure accommodates 11 assets (10 risky assets + 1 risk-free asset).

## 5 EMPIRICAL TESTS

### 5.1 Data Selection, Preprocessing and Assumptions

This study constructs an investment portfolio using randomly selected constituent stocks from the CSI300 index for empirical analysis. The portfolio consists of 1 risk-free asset (cash) and 10 risky assets. The data is obtained from the Wind database's daily trading records, with all prices forward-adjusted. The study restricts trading to one transaction per stock per day.

This research implements random portfolio selection, departing from traditional investment theory's selective strategies based on liquidity, diversification, and other factors. The methodology stems from the premise that an effective DRL model should demonstrate adaptability across diverse portfolios, beyond carefully selected asset combinations. Superior backtesting performance of randomly selected portfolios provides empirical validation of the DRL model's decision-making capabilities and generalization effectiveness. This approach exemplifies DRL's fundamental advantage as a data-driven model: autonomous adaptation to market environments without manual asset screening procedures.

For asset selection, the study applies a single temporal criterion: assets must have been listed before December 31, 2012. This requirement reflects the data-driven nature of DRL models, which require substantial historical data for training. Extended listing histories provide comprehensive trading data, enabling enhanced market feature learning. In the random asset selection process, preference is given to stocks exhibiting overall upward price trends historically, as these patterns provide more meaningful training signals for long-only DRL models. The study assumes sufficient liquidity for risky assets, immediate transaction execution, and negligible market impact from trading activities.

### 5.2 Performance Metrics, Backtesting Period and Comparative Optimization Models

Following Zhang et al.[17, 40], this study incorporates performance metrics encompassing annualized average return E(R), annualized volatility Std(R), annualized Sharpe ratio (Sharpe), annualized Sortino ratio (Sortino), maximum drawdown (MDD), Calmar ratio (Calmar), percentage of positive returns (%of+Ret), and average profit-loss ratio (Ave P/Ave L).

The research implements a six-month backtesting period to evaluate model optimization effectiveness. With 252 trading days per year, the selected 128 trading days represent approximately six months of trading time. To address overfitting concerns, the methodology adapts Wassname's [50] open-source github implementation, sampling 128 consecutive trading days randomly from the complete dataset for each training episode. The six-month backtesting period selection aligns with this sampling framework.

Figure 4 depicts the training and testing sets. Price data undergoes standardization, with each asset's price normalized by its final trading day opening price, enabling clear trend visualization across portfolio assets with diverse price levels.

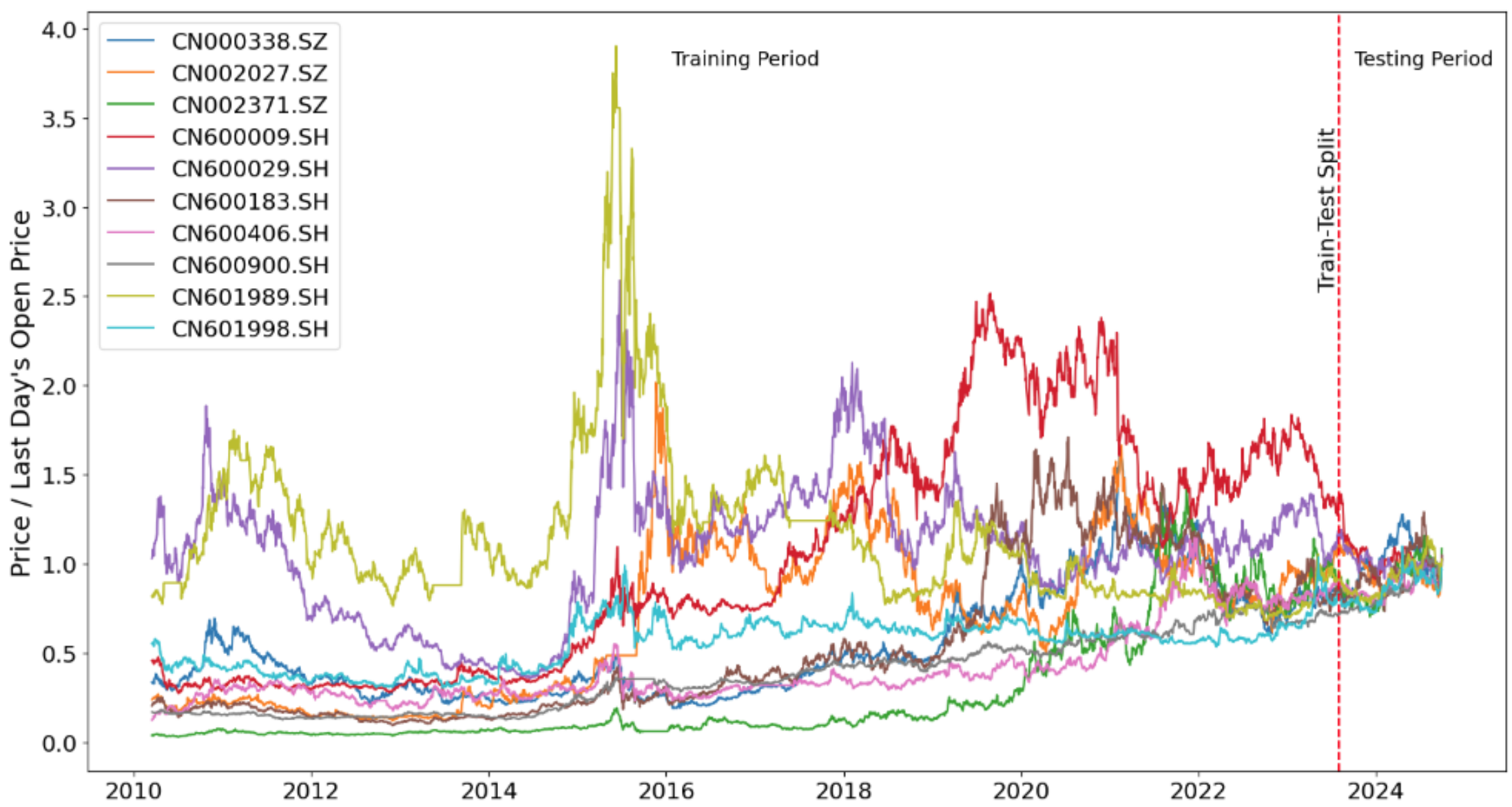

Fig4 Normalized Price Trends with Train-Test Split

The backtesting period encompasses exclusively out-of-sample data, independent from the training dataset. The agent (i.e., the investor) processes these data exclusively during backtesting, without prior exposure to future price movements. Table 1 delineates the specific time periods for training and backtesting data:

**Table 1 Data ranges for training and backtesting**

| Asset | Training Period | Testing Period |
|---|---|---|
| Stock Portfolio | 03/17/2010 - 08/02/2023 | 08/07/2023 - 02/20/2024 |

The comparative analysis framework incorporates multiple established optimization models benchmarked against the DRL model. These models are implemented through the Riskfolio-lib asset optimization package, maintaining default configurations across all comparative models with asset returns derived from closing prices. The optimization framework encompasses: Classic Mean Variance (MV), Conditional Value at Risk (CVaR), Entropic Value at Risk (EVaR), Risk Parity (RP), Hierarchical Risk Parity (HRP), Hierarchical Equal Risk Contribution (HERC), and Nested Clustered Optimization (NCO). While these models support various objective functions including risk minimization (MinRisk), Sharpe ratio maximization (Sharpe), utility function maximization (Utility), and net asset value maximization (MaxRet), the comparative analysis focuses exclusively on risk minimization and Sharpe ratio maximization, given the subjective nature of utility functions and the empirical underperformance of utility and net asset value maximization strategies.

The historical data window selection for comparative models adheres to the methodological framework established by the original authors of EVaR[51] and HRP[12], employing 4-year and 1-year periods respectively. With an annual trading calendar of 252 days, the 4-year period encompasses 1,008 trading days (252 * 4). Given that these quantitative optimization models conceptualize asset weight modifications as static processes, disregarding weight dynamics in continuous trading, the study implements a rolling window methodology for weight prediction. Specifically, weight predictions for September 1, 2021, utilize historical data from the preceding 4 or 1 years through August 31, 2021, with this process continuing throughout the backtesting period. Transaction costs emanating from weight adjustments are computed using formula 10, maintaining consistency with the transaction cost parameters established in the DRL model.

5.3 Training Results and Robustness Testing

Deep Reinforcement Learning (DRL) constitutes a novel sequential statistical decision-making methodology that leverages neural networks to model and estimate conditional probability distributions and expected returns across the state-action space. At each timestep, the agent executes online statistical inference based on current observations while optimizing its decision strategy through systematic exploration and experience accumulation, implementing an iterative statistical learning process designed to maximize expected cumulative rewards. This methodology integrates the function approximation capabilities of deep learning with the sequential decision-making framework of reinforcement learning, establishing an end-to-end statistical modeling and optimization approach.

While traditional econometric testing methods based on linear assumptions struggle to effectively evaluate the statistical significance of DRL models, the convergence of training rewards provides a more appropriate evaluation criterion. The convergence behavior indicates the agent's ability to consistently generate profits in historical market environments, serving as a necessary condition for model stability and robustness. This necessity stems from DRL's non-linear nature and its adaptive learning mechanisms - a converged reward function demonstrates the model has learned stable patterns rather than overfitting to market noise. Furthermore, reward convergence implies the agent has developed a generalizable strategy that maintains consistent performance across various market conditions within the training distribution.

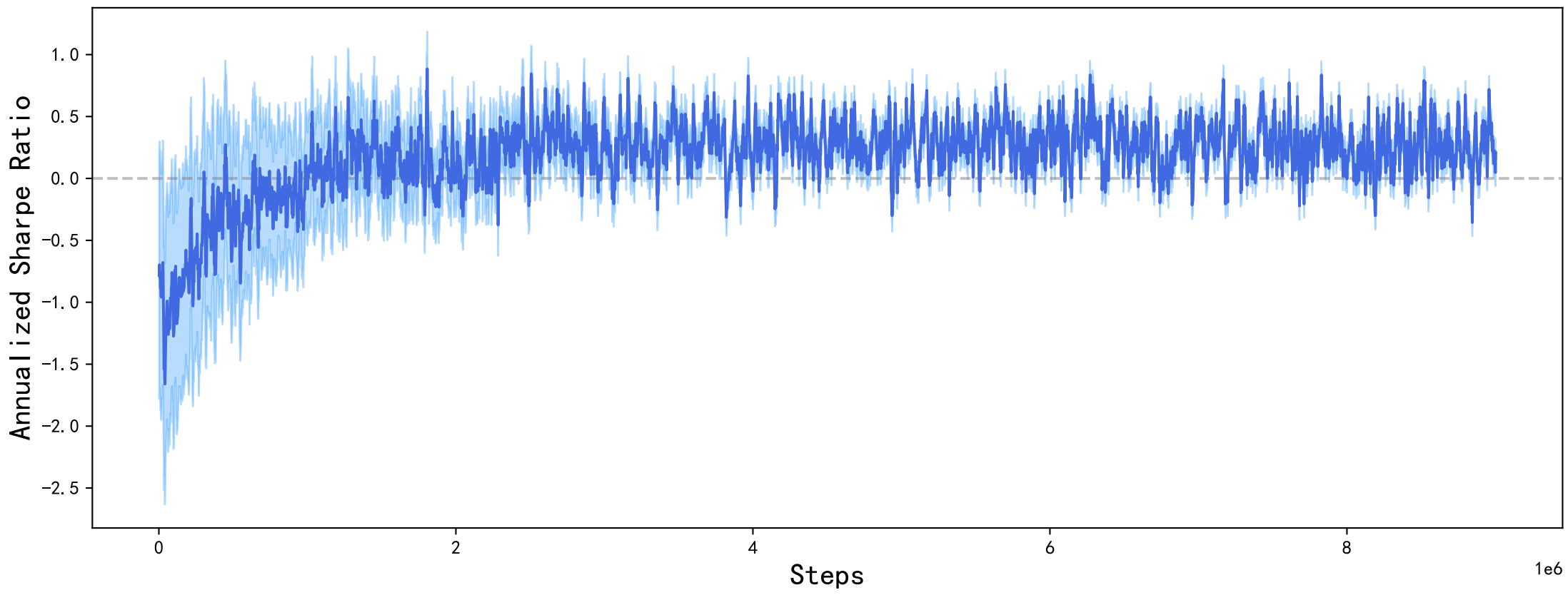

Figure 5 Training rewards

As demonstrated in Figure 5, the model underwent training for 9 million steps. The agent's acquired rewards exhibit a positive correlation with the progression of training steps, indicating systematic improvement in the training process and agent performance. Throughout the training period, the reward values demonstrate convergence, with the annualized Sharpe ratio stabilizing within the range of -0.3 to 0.8, and the predominant portion of training reward values maintaining convergence above zero. These results indicate that the agent demonstrates consistent return generation capability within the known environment, supporting the robustness of the trained model.

### 5.4 Backtesting results

#### 5.4.1 Portfolio Value, Asset Allocation and Trading Costs

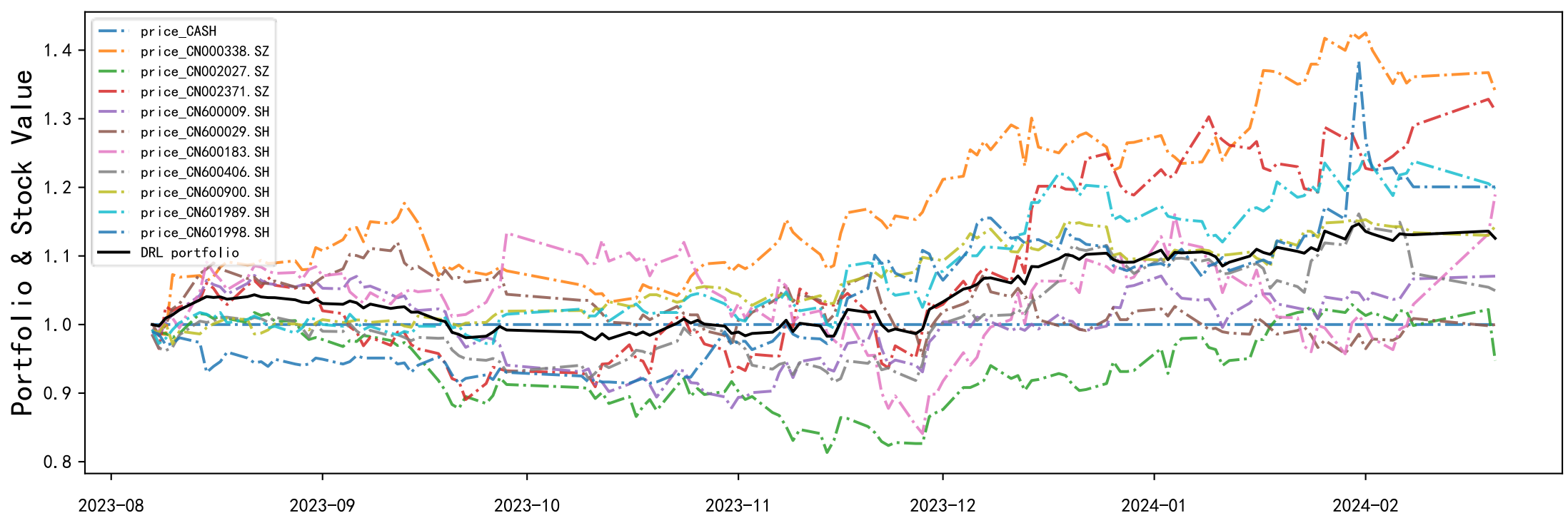

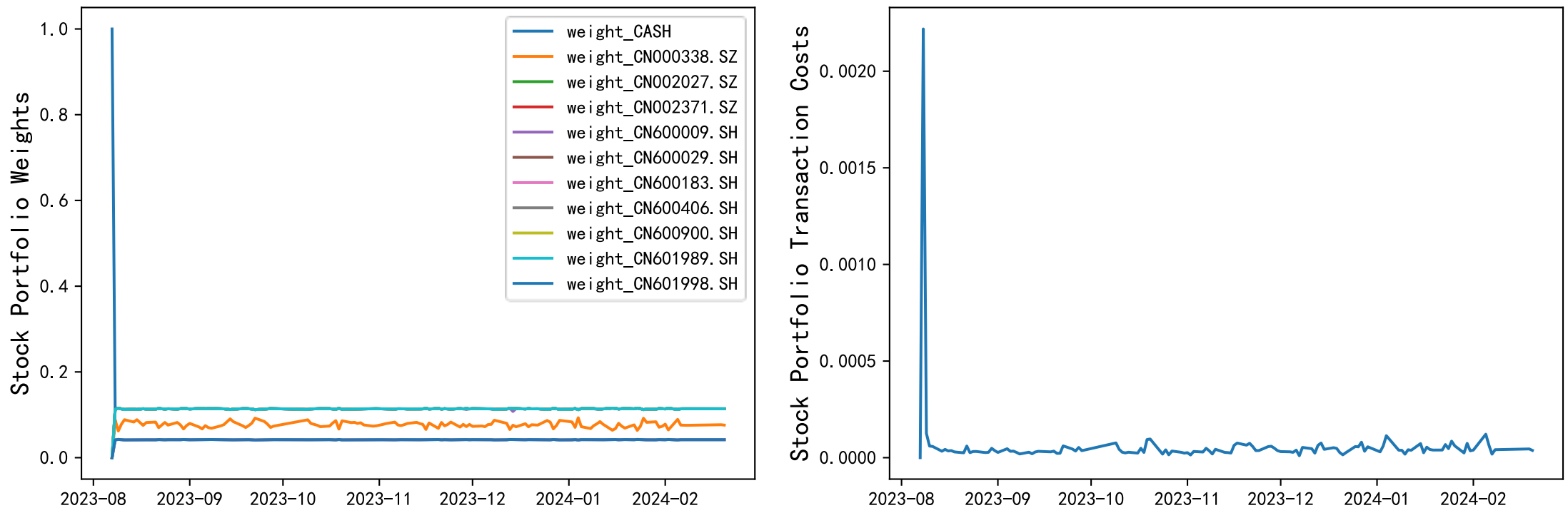

Figure 6 DRL Portfolio Value, Asset Allocation and Trading Costs (2023.08-2024.02)

The upper panel of Figure 6 presents the relative prices of assets (calculated according to formula 4). The relative price can be viewed as a standardization method to adjust the portfolio assets' prices to the same scale. The lower panels display the portfolio's asset weights and transaction costs respectively.

The DRL portfolio value demonstrated a consistent upward trajectory throughout the backtesting period, appreciating from an initial value of 1.0 to 1.1256, generating a total return of 12.56%. Two significant upward movements materialized during mid-November to December 2023 and late January 2024. Despite experiencing a drawdown in mid-September 2023, where the portfolio value temporarily declined to approximately 0.98, the drawdown magnitude remained modest, followed by robust recovery momentum.

Regarding weight allocation, DRL implemented a robust asset allocation strategy. The portfolio comprises cash and 10 stocks, with initial allocations approximating uniform distribution at 0.0909 per asset. Throughout the trading period, DRL maintained consistent weight adjustment patterns, with weight standard deviation stabilizing between 0.031 and 0.033, demonstrating effective diversification characteristics. The cash position fluctuated within a narrow range of 0.11 to 0.12, ensuring adequate portfolio liquidity. In terms of stock weights, CN002027.SZ and CN002371.SZ exhibited relatively active weight adjustments, while CN600029.SH and CN600183.SH maintained consistently lower allocation ratios.

Throughout the backtesting period, the strategy exhibited efficient transaction cost management. Substantial transaction costs (approximately 0.22%) were incurred solely during initial capital allocation, with subsequent transaction cost rates maintaining minimal levels between 0.002% and 0.01% on most trading days. This performance indicates the implementation of a measured trading approach in asset allocation adjustments, effectively mitigating the impact of transaction costs on portfolio returns.

The PPO algorithm-based portfolio demonstrated favorable risk-return characteristics, generating positive investment returns through maintained diversification and dynamic weight adjustments, while effectively managing downside risk. These results suggest significant potential for deep reinforcement learning applications in portfolio management.

5.4.2 Performance Comparison of Stock Optimization Models

Table 2 presents the performance comparison of various stock optimization models. Models are designated according to the convention "model type-optimization objective-data window", illustrated as follows:

- MV-MinRisk: Mean-variance model, with risk minimization as the optimization objective, employing a 1-year historical data rolling window.
- CVaR-Sharpe-4yr: Conditional Value-at-Risk model, with Sharpe ratio maximization as the optimization objective, employing a 4-year historical data rolling window.

Additional models follow this naming convention.

Table 2 Results of various optimization models for stocks

| | **E(R)** | **Std(R)** | **Sharpe** | **Sortino** | MDD | Calmar | **%of+Ret** | $\frac{\text{Ave. P}}{\text{Ave. L}}$ |
|---|---|---|---|---|---|---|---|---|
| DRL | 0.1956 | 0.1258 | 1.5550 | 2.9567 | 5.85% | 3.3395 | 0.4728 | 1.4204 |
| MV-MinRisk | 0.0892 | 0.1158 | 0.7707 | 1.2357 | 5.88% | 1.5175 | 0.5156 | 1.0641 |
| MV-MinRisk-4yr | 0.1016 | 0.1195 | 0.8501 | 1.4115 | 6.37% | 1.5940 | 0.4843 | 1.2285 |
| MV-Sharpe | 0.0738 | 0.1584 | 0.4662 | 0.7835 | 8.20% | 0.8998 | 0.4609 | 1.2726 |
| MV-Sharpe-4yr | 0.0283 | 0.1288 | 0.2199 | 0.3920 | 9.41% | 0.3011 | 0.4687 | 1.1738 |
| CVaR-MinRisk | 0.0817 | 0.1155 | 0.7073 | 1.1312 | 6.26% | 1.3039 | 0.5156 | 1.0566 |
| CVaR-MinRisk-4yr | 0.1508 | 0.1227 | 1.2290 | 2.0967 | 5.38% | 2.8041 | 0.4843 | 1.3101 |
| CVaR-Sharpe | 0.0326 | 0.1772 | 0.1840 | 0.2844 | 9.01% | 0.3618 | 0.4531 | 1.2492 |
| CVaR-Sharpe-4yr | 0.0049 | 0.1349 | 0.0367 | 0.0638 | 10.80% | 0.0459 | 0.4609 | 1.1763 |
| EVaR-MinRisk | 0.0350 | 0.1187 | 0.2953 | 0.4870 | 7.92% | 0.4423 | 0.4843 | 1.1175 |
| EVaR-MinRisk-4yr | 0.1450 | 0.1297 | 1.1172 | 1.9260 | 6.13% | 2.3635 | 0.5312 | 1.0660 |
| EVaR-Sharpe | -0.0007 | 0.1742 | -0.0042 | -0.0059 | 9.16% | -0.0080 | 0.4218 | 1.3692 |
| EVaR-Sharpe-4yr | 0.0339 | 0.1392 | 0.2435 | 0.4341 | 9.40% | 0.3607 | 0.4609 | 1.2154 |
| RP-MinRisk | -0.0231 | 0.1342 | -0.1721 | -0.2736 | 9.38% | -0.2461 | 0.4453 | 1.2106 |
| RP-MinRisk-4yr | -0.0154 | 0.1318 | -0.1169 | -0.1833 | 9.20% | -0.1675 | 0.4375 | 1.2606 |
| RP-Sharpe | -0.0231 | 0.1342 | -0.1721 | -0.2736 | 9.38% | -0.2461 | 0.4453 | 1.2106 |
| RP-Sharpe-4yr | -0.0154 | 0.1318 | -0.1169 | -0.1833 | 9.20% | -0.1675 | 0.4375 | 1.2606 |
| HRP | 0.0359 | 0.1241 | 0.2893 | 0.4553 | 7.19% | 0.4988 | 0.4609 | 1.2269 |
| HRP-4yr | 0.03100 | 0.1243 | 0.2492 | 0.3946 | 7.98% | 0.3880 | 0.4843 | 1.1106 |
| HERC | -0.1007 | 0.1372 | -0.7339 | -1.1000 | 11.86% | -0.8488 | 0.4375 | 1.1361 |
| HERC-4yr | 0.0288 | 0.1249 | 0.2310 | 0.3630 | 7.99% | 0.3610 | 0.4687 | 1.1787 |
| NCO-MinRisk | 0.0768 | 0.1155 | 0.6652 | 1.0334 | 6.00% | 1.2806 | 0.5 | 1.1162 |
| NCO-MinRisk-4yr | 0.1064 | 0.1188 | 0.8961 | 1.4863 | 5.99% | 1.7750 | 0.4843 | 1.2404 |
| NCO-Sharpe | 0.0529 | 0.1600 | 0.3310 | 0.5574 | 8.07% | 0.6557 | 0.4531 | 1.2816 |
| NCO-Sharpe -4yr | 0.0155 | 0.1265 | 0.1228 | 0.2186 | 9.89% | 0.1571 | 0.4609 | 1.1926 |

Adopting Ernest P. Chan[52]'s approach of utilizing the Sharpe ratio as the primary metric, Table 2 demonstrates that the Deep Reinforcement Learning (DRL) model exhibits significant advantages in portfolio optimization. The model achieved the highest annualized average return of 19.56%, maintained moderate volatility (12.58%), and notably outperformed other models with a Sharpe ratio of 1.5550, demonstrating superior risk-adjusted returns. The model attained the highest Sortino ratio (2.9567), validating its exceptional performance when considering downside risk. In terms of risk management, the DRL model exhibited a maximum drawdown of merely 5.85%, approaching optimal levels, while achieving the highest Calmar ratio (3.3395), indicating superior downside risk management capabilities.

Among traditional optimization models, CVaR-MinRisk-4yr demonstrated optimal performance, generating a 15.08% annualized return and the lowest maximum drawdown (5.38%). This model achieved a Sharpe ratio of 1.2290, which, although lower than the DRL model, represented the highest among traditional models. The superior performance may be attributed to its 4-year lookback period design, facilitating more stable historical data analysis. Notably, models implementing the MinRisk strategy consistently outperformed their Sharpe strategy counterparts, suggesting superior effectiveness of risk minimization compared to risk-adjusted return maximization in the current market environment.

Significantly, Risk Parity (RP) series models and Hierarchical Equal Risk Contribution (HERC) strategy exhibited suboptimal performance. RP models consistently generated negative returns and Sharpe ratios, while HERC models recorded a maximum negative return of -10.07% and maximum drawdown of 11.86%, indicating that excessive dependence on historical correlations or simplified risk allocation methodologies may insufficiently address market dynamics.

From a trading efficiency perspective, the DRL model achieved superior performance with an investment win rate of 47.28% and optimal average gain-loss ratio of 1.4204, demonstrating proficiency in both market opportunity capture and loss mitigation. Comprehensively, whether assessed through the primary Sharpe ratio metric or alternative risk-adjusted return indicators, the DRL model effectively integrated return generation capabilities with risk management efficiency, surpassing traditional methodologies across all dimensions while exhibiting adaptability and robustness in complex market environments. These findings validate the application potential of deep reinforcement learning in finance, particularly its efficacy in portfolio management requiring dynamic decision-making and multi-objective optimization.

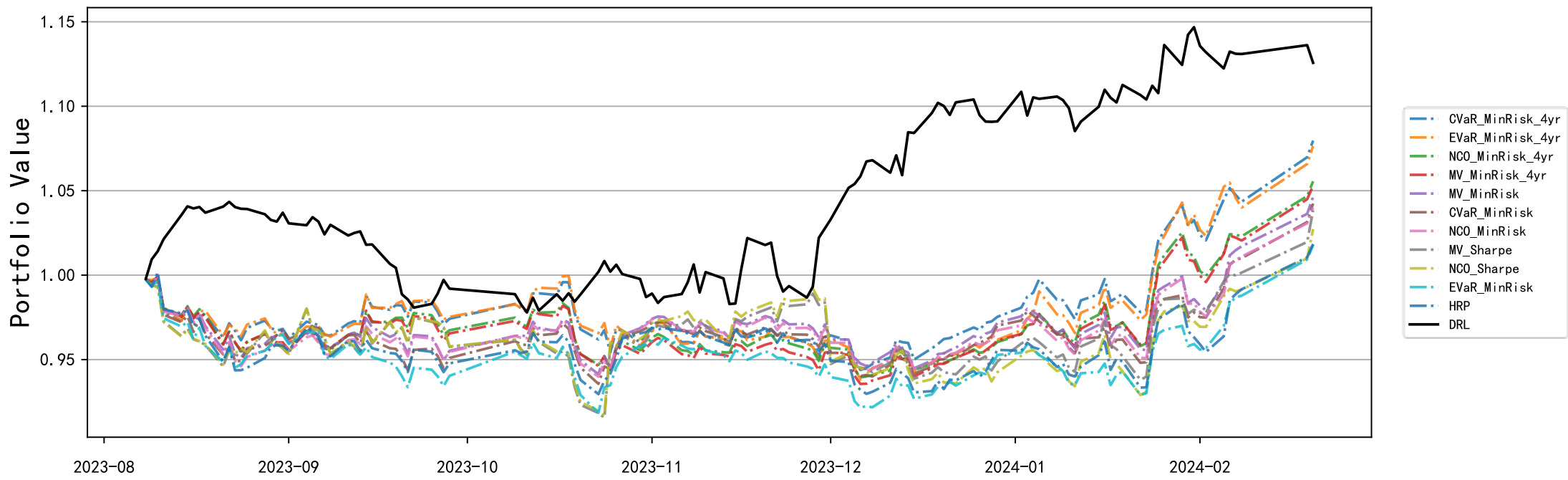

Figure 7 Performance Comparison between DRL and Other Optimization Models

To facilitate quantitative comparison, this study examines the top 11 traditional optimization models ranked by annualized Sharpe ratio in comparison with the DRL model (as shown in Figure 7). The empirical results indicate that throughout the backtesting period, the DRL model exhibits superior performance relative to traditional optimization models, as evidenced by the following findings:

i. Return metrics: The DRL model consistently outperforms traditional optimization models across the backtesting period, generating substantial positive returns. Within the traditional optimization framework, CVaR-MinRisk-4yr, EVaR-MinRisk-4yr, NCO-MinRisk-4yr, and MV-MinRisk-4yr yield the highest performance metrics, yet remain below those of the DRL model.

ii. Strategic characteristics: Traditional optimization models demonstrate strong homogeneity, with portfolio value trajectories following virtually identical patterns post-optimization. In contrast, the DRL-optimized portfolio value trajectories display distinct patterns from those of traditional optimization approaches. This suggests that the data-driven DRL framework demonstrates enhanced capability in capturing asset price dynamics, leading to more efficient asset weight allocation.

Empirical evidence from extensive experimental studies indicates that the DRL model exhibits substantial efficacy in portfolio optimization of CSI300 constituent stocks. This performance can be attributed to two key factors: firstly, the backtesting period coincided with an upward trajectory of CSI300 constituent stocks, creating favorable conditions for long-strategy validation; secondly, CSI300 constituent stocks maintain a stable investor composition characterized by a higher concentration of institutional investors, resulting in more systematic investment patterns. Relative to small and medium-cap segments, the price and trading data of CSI300 constituent stocks demonstrate enhanced reliability in reflecting market fundamentals and investor sentiment, thereby establishing a more robust learning environment for the DRL model. These structural characteristics facilitate improved learning and market adaptation capabilities of the DRL model, leading to enhanced performance in CSI300 constituent stock portfolio optimization.

## 6 CONCLUSIONS

Traditional financial econometric optimization models demonstrate inherent limitations in portfolio asset allocation due to their static frameworks and homogeneous characteristics. While these approaches facilitate theoretical analysis, they struggle to capture the dynamic evolution of asset weights in real trading environments and show insufficient adaptability to market volatility.

This research advances the application of Deep Reinforcement Learning (DRL) in portfolio optimization through two key innovations. First, we introduce a novel Sharpe ratio reward function specifically designed for Actor-Critic DRL algorithms, achieving stable convergence during training and consistently positive Sharpe ratios in practice. Second, we develop a comprehensive DRL methodology integrating VGG-based neural networks for multi-dimensional financial data processing with random sampling strategies, effectively enhancing model generalization while minimizing overfitting risk. Empirical results demonstrate the superior performance of our DRL framework compared to traditional optimization models across multiple metrics. The DRL model consistently achieves higher returns than leading traditional approaches including CVaR-MinRisk-4yr and EVaR-MinRisk-4yr, while exhibiting distinct portfolio value trajectories that indicate enhanced capability in capturing asset price dynamics.

With the continuous advancement of deep reinforcement learning (DRL) theory, its applications in finance are demonstrating significant potential. DRL optimization methods integrate knowledge across multiple disciplines including machine learning, finance, and statistics. This interdisciplinary convergence provides novel research perspectives and methodological foundations for portfolio optimization. We observe that researchers frequently underemphasize the importance of DRL environment modeling when applying DRL models to portfolio optimization. The modeling of DRL environments encompasses the operational logic of the entire model. The innovative average Sharpe ratio reward function introduced in this paper is implemented within the environment modeling framework, and the quality of environment modeling directly impacts model performance. Furthermore, financial data inherently contains substantial noise. The effective utilization of this data for environment modeling in DRL and extraction of valuable trading signals is crucial for enhancing model performance.

Looking forward, research directions should focus not only on optimizing DRL algorithms but also on deeply exploring innovative approaches to environment modeling. As a core component of the DRL framework, environment modeling directly determines the model's capacity to perceive and adapt to market dynamics, which is of decisive importance for increasing the practical application value of DRL in portfolio optimization.

# APPENDIX

## Results of other optimization models for Stocks

Column 1: Portfolio value; Column 2: Asset weights; Column 3: Transaction cost rate

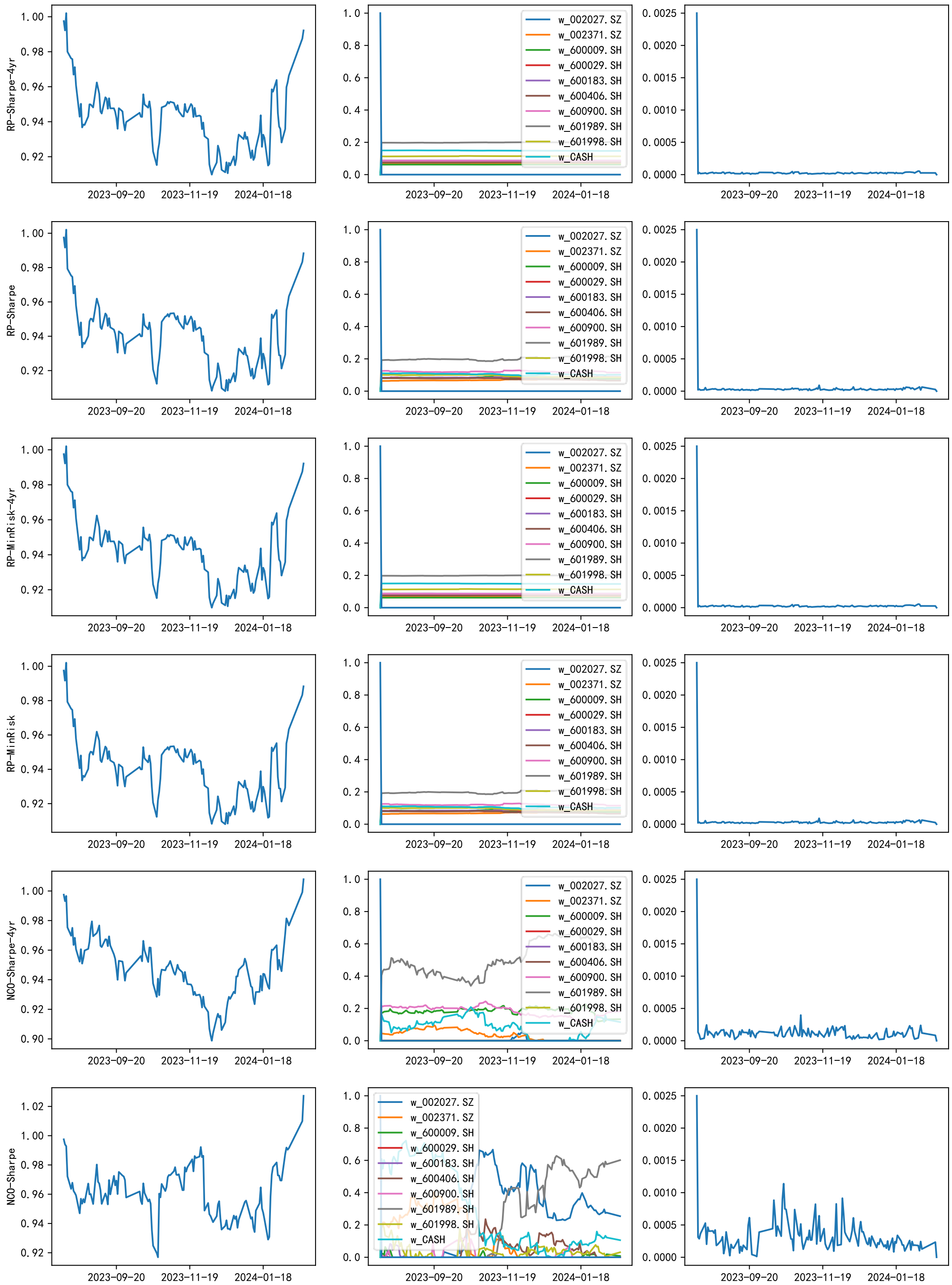

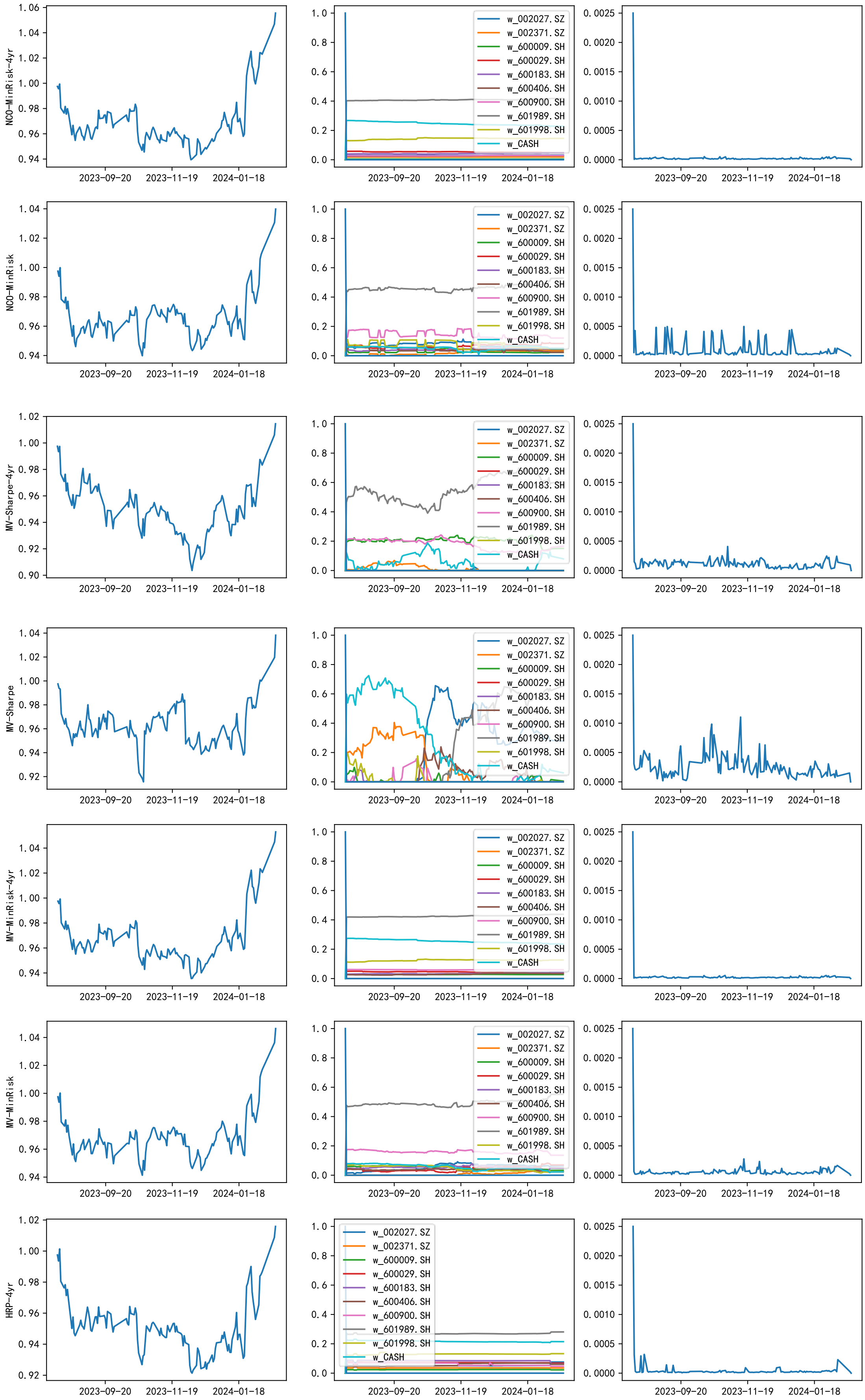
NCO-MinRisk-4yr
NCO-MinRisk
MV-Sharpe-4yr
MV-Sharpe
MV-MinRisk-4yr
MV-MinRisk
HRP-4yr
2023-09-20
2023-11-19
2024-01-18
w_002027.SZ
w_002371.SZ
w_600009.SH
w_600029.SH
w_600183.SH
w_600406.SH
w_600900.SH
w_601989.SH
w_601998.SH
w_CASH

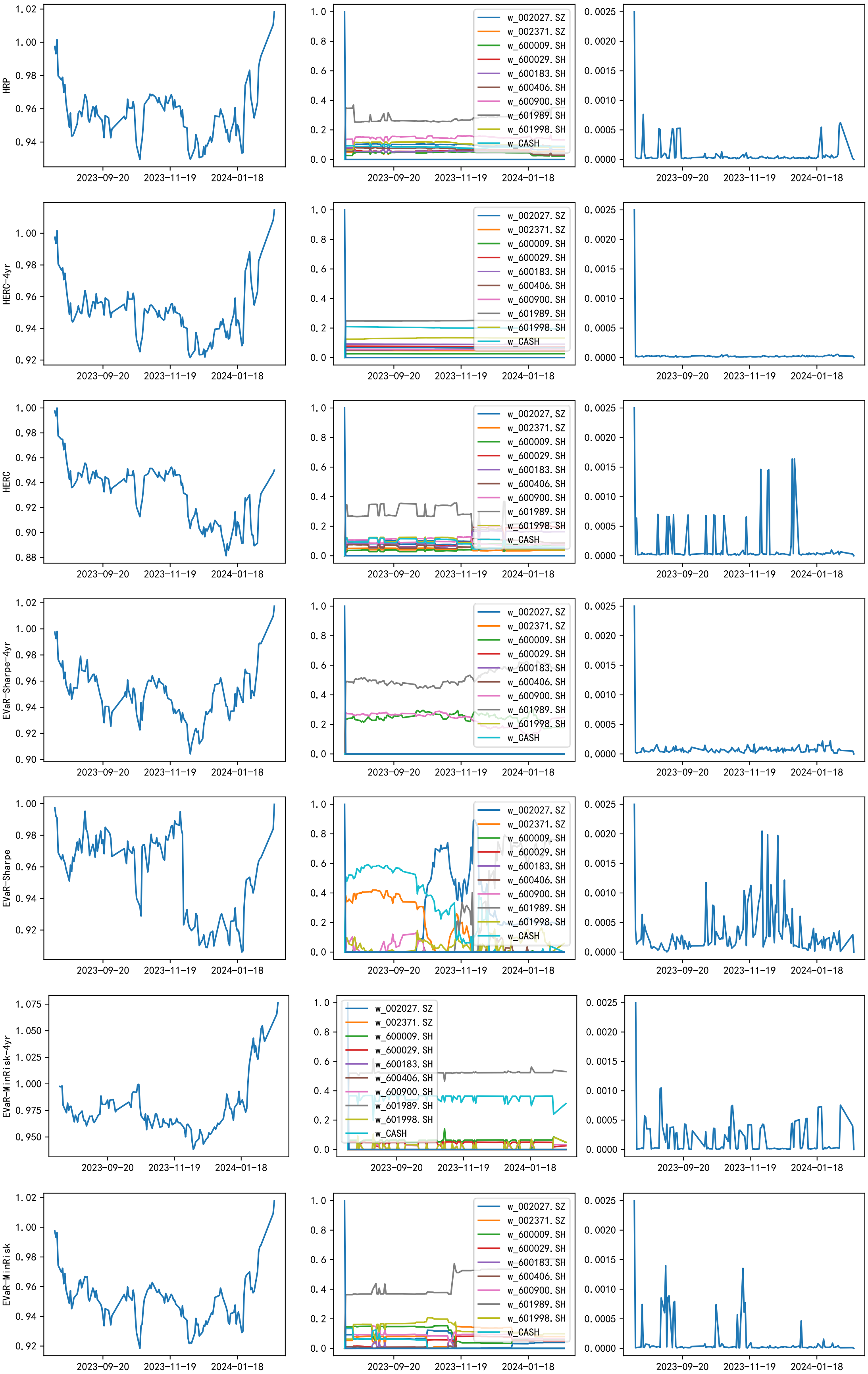
HRP
HERC-4yr
HERC
EVaR-Sharpe-4yr
EVaR-Sharpe
EVaR-MinRisk-4yr
EVaR-MinRisk
w_002027.SZ
w_002371.SZ
w_600009.SH
w_600029.SH
w_600183.SH
w_600406.SH
w_600900.SH
w_601989.SH
w_601998.SH
w_CASH
2023-09-20
2023-11-19
2024-01-18

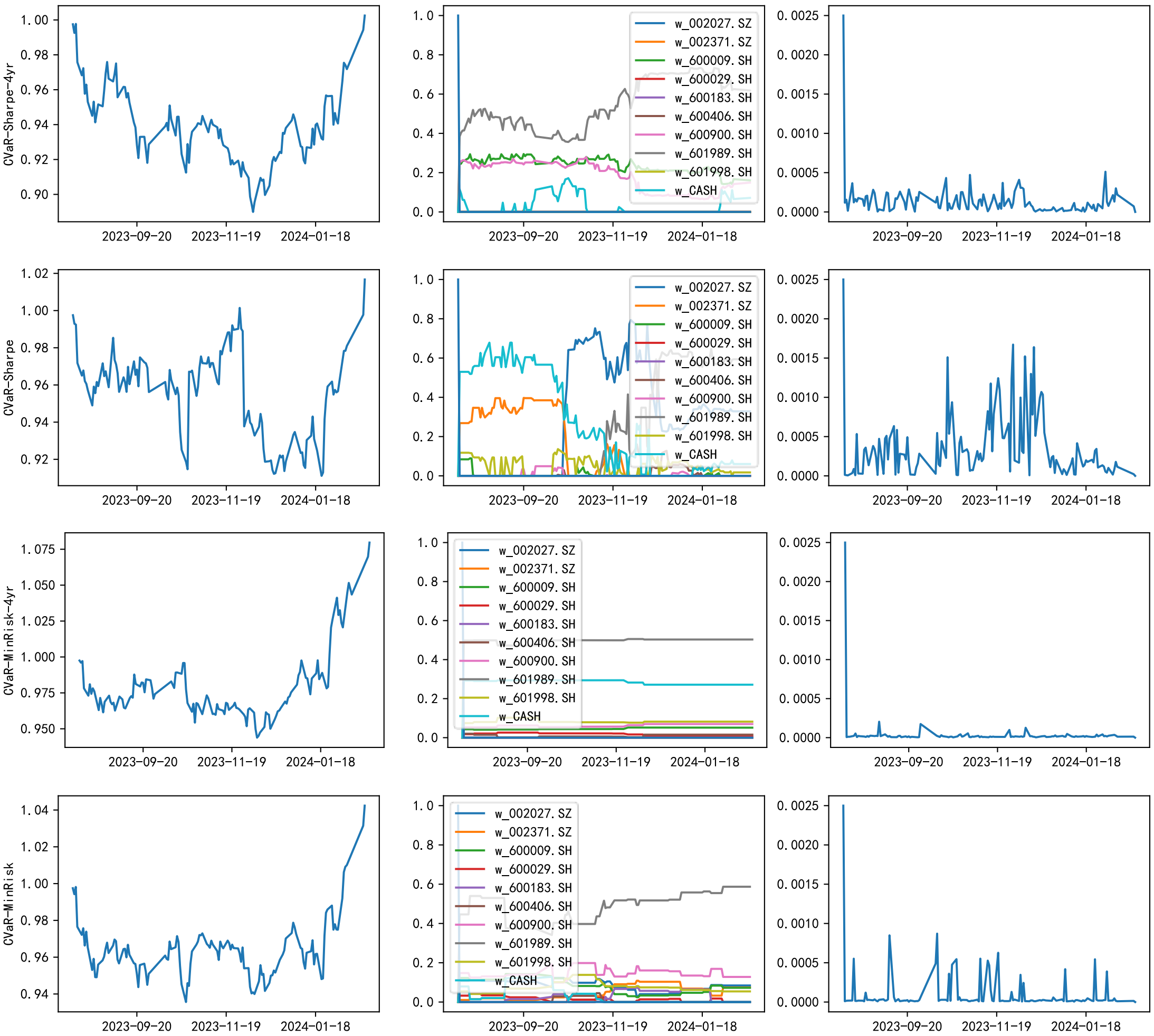

CVaR-Sharpe-4yr
CVaR-Sharpe
CVaR-MinRisk-4yr
CVaR-MinRisk
w_002027. SZ
w_002371. SZ
w_600009. SH
w_600029. SH
w_600183. SH
w_600406. SH
w_600900. SH
w_601989. SH
w_601998. SH
w_CASH
2023-09-20
2023-11-19
2024-01-18